\documentclass[twocolumn,showpacs,preprintnumbers,amsmath,amssymb,superscriptaddress,bibnotes]{revtex4}

\usepackage{graphicx}
\usepackage{dcolumn}
\usepackage{bm}

\usepackage{soul}

\usepackage{epstopdf}
\usepackage{color}

\begin{document}	
\preprint{APS/123-QED}
\title{Heterogeneity induces emergent functional networks for synchronization}

\author{Francesco Scafuti}
\affiliation{Department of Electrical Engineering and Information Technology, University of Naples Federico II, Naples, Italy}
\author{Takaaki Aoki}
\affiliation{Faculty of Education, Kagawa University, Takamatsu, Japan}
\author{Mario di Bernardo}
\email{mario.dibernardo@unina.it}
\affiliation{Department of Electrical Engineering and Information Technology, University of Naples Federico II, Naples, Italy}
\affiliation{Department of Engineering Mathematics and Bristol Centre for Complexity Sciences, University of Bristol, Bristol, UK}

\date{\today}

\begin{abstract}
We study the evolution of heterogeneous networks of oscillators subject to a state-dependent interconnection rule. We find that heterogeneity in the node dynamics is key in organizing the architecture of the functional emerging networks. We demonstrate that increasing heterogeneity among the nodes in state-dependent networks of phase oscillators causes a differentiation in the activation probabilities of the links. This, in turn, yields the formation of  hubs associated to nodes with larger distances from the average frequency of the ensemble. Our generic local evolutionary strategy can be used to solve a wide range of synchronization and control problems.
\end{abstract}
\maketitle

\section{Introduction}
Evolution is a fundamental force driving the organization and structure of natural systems. It is based on two key ingredients: variation and natural selection \cite{Darw:51}. The first ensures the necessary mutation and recombination generating new species while the second determines the survival of the fittest to perform a certain function.
Networks in Nature have been subject to the same powerful mechanisms that ultimately determined their structure, properties and functionality. The resulting networks have heterogenous topological structures, which researchers have been interested in together with their effects on dynamical processes \cite{TaLu:14}. Examples include epidemic spreading, opinion formation, and synchronization \cite{Boccaletti2006175,barrat2008dynamical}.  Often there is also heterogeneity in the nodes of a network. For example, in social networks, individuals have different personalities, which will have great impacts on their social relationships; or, in manufacturing, industrial products are slightly different from one other, affecting their impact and market shares.
The relationship between the heterogeneity of the nodes and the structural properties of a network is little understood, particularly when the network evolution is state-dependent.

Here we suggest that {\em heterogeneity in the nodes is a driving force behind the evolution of the network structure that determines its properties and function}.
To test this ansatz we take as a representative example the problem of evolving the network structure to achieve synchronization of coupled oscillators. This is one of the best understood and most widely studied type of collective behavior on networks \cite{KuramotoChemicalOsci, Strogatz20001,KuramotoReview,ArGu:08,barrat2008dynamical}.

So far, optimal network structures for synchronization have been studied mainly by using Monte Carlo methods \cite{PhysRevLett.95.188701,Chaos2008,PhysRevE.81.056204,PhysRevE.81.056212} or gradient-based learning strategies \cite{Tanaka2008b,PhysRevLett.113.144101}. These are based on the use of some objective function for synchronization (as for example the order parameter) which is used to find the optimal network whose structural properties  are then surveyed.
The Monte Carlo approach is a generic and powerful strategy  but it is typically time-consuming, and increasingly cumbersome to apply to large-scale networks. Gradient-based methods assume some constraints to derive the evolution rule of the coupling strengths and the rules are often not local, in the sense that  some global information on the entire network is used. Also, it has been shown that adaptive networks can yield the emergence of modular and scale-free structures, while enhancing synchronization \cite{GuAm:11}.

In this paper, we propose the use of an evolutionary strategy to find a functional structure for synchronization in a network of heterogeneous oscillators. In so doing we will show that heterogeneity in the nodes is instrumental in determining the properties of the resulting network. The goal of the strategy is to identify, over all possible unweighted network configurations,  the structure with a minimal number of links, which  guarantees frequency synchronization of its nodes.
While the fundamental aim of our study is similar to that of the literature \cite{PhysRevLett.95.188701,Chaos2008,PhysRevE.81.056204,PhysRevE.81.056212,Tanaka2008b,PhysRevLett.113.144101,GuAm:11}, the approach we propose is completely different. Indeed, our strategy uses adaptive schemes which are completely local and do not rely on any global synchronization measure. Moreover such schemes are deployed in a novel evolutionary manner.

\section{Problem statement}
We start by considering a network of general nonlinear coupled oscillators
\begin{equation}\label{equ:genoscill}
\dot{\bm{x}}_n=\bm{f}_n(x_n)+c \sum_{m=1}^{N} k_{nm} \bm{g}(\bm{x}_m, \bm{x}_n),
\end{equation}
where $\bm{x}_n \in \mathbb{R}^p$ is the $p$-dimensional state of the $n$-th oscillator, $\bm{f}_n$ denotes its dynamics (note that oscillators can be slightly different from each other due to both parameters and model mismatches), $\bm{g}$ is a generic coupling function and $k_{nm}$ are time-varying coupling gains determining the strength of the coupling between neighboring oscillators.
\begin{figure}[tb]
\includegraphics{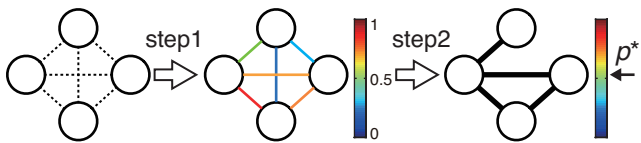}
\caption{Schematic description of the evolutionary edge-snapping strategy. Step 1 (variation): computation of  link activation probabilities by running the edge-snapping strategy from many different random initial conditions. Step 2 (selection): selection of those links whose activation probability is above some threshold value $p^*$. }
\label{fig:1}
\end{figure}

We model the evolutionary pressures to reach synchronization  by considering state-dependent second-order nonlinear dynamics for the gains dependent upon a double well potential $V(x) = b x^2 (x-1)^2$. The gain dynamics are given by
\begin{equation}\label{equ:genedgesn}
\ddot{k}_{nm}+d \ \dot{k}_{nm}+\frac{\partial V(k_{nm})}{\partial k_{nm}}=h(\| \bm{x}_m- \bm{x}_n\|),
\end{equation}
in which $h(\| \bm{x}_m- \bm{x}_n\|)$ is a generic increasing function such that $h(0)=0.$
Note that this is a very general adaptive network equation relying on a decentralized, local, state-dependent interconnection rule. This system can be systematically reduced, under a standard technique \cite{KuramotoChemicalOsci}, to the network of adaptively coupled phase oscillators:
\begin{align}
&\dot{\theta}_n =\omega_n+\frac{1}{N}\sum_{m=1}^{N}k_{nm} \Gamma(\theta_m-\theta_n), \label{equ:genphoscill} \\
&\ddot{k}_{nm}+d \ \dot{k}_{nm}+\frac{\partial V(k_{nm})}{\partial k_{nm}} = h(\|\theta_m - \theta_n \|), \label{equ:genedgesnph}
\end{align}
in which $\theta_n$ is the phase of the $n$-th generic oscillator, $\Gamma(\theta_m-\theta_n)$ is a generic $2\pi$-periodic function.
We set the overall coupling strength $K$ to a unitary value, since it can be absorbed into a parameter defining the heterogeneity of the natural frequencies by rescaling time, i.e. by setting $\tau=K t$.
In this paper  we analyze, for the sake of clarity, the simplest case 
\begin{align}
\Gamma(\theta_m-\theta_n) &=\sin(\theta_m-\theta_n), \label{equ:Gamma}\\
h(\|\theta_m - \theta_n \|) &=\alpha\left[1-\frac{1}{2} |e^{i \theta_n}+e^{i \theta_m}| \right]. \label{equ:unm}
\end{align}

The effectiveness of edge snapping strategies to achieve synchronization has been discussed in \cite{EdgeSnapping} and further details are given in Appendix A.
Under such a forcing the dynamics of $k_{nm}$ (starting from zero initial conditions $k_{nm}(0)=0$ and $\dot{k}_{nm}(0)=0$), will either converge towards 0 (link is not present) or towards 1 (link is activated)

The differences in the natural frequencies of the oscillators originate from the heterogeneity of the node dynamics $\bm{f}_n$ in weakly coupled nonlinear oscillators \cite{KuramotoChemicalOsci}.
In what follows, these natural frequencies are selected deterministically from a Gaussian distribution with zero mean and standard deviation equal to $\sigma$. Therefore, the parameter $\sigma$ can be used to ``tune'' the level of heterogeneity among nodes.

We note here that when the number of nodes is not so large, such as $N = 6$ or $7$, the natural frequencies sampled from a distribution can be biased.
To avoid the effect of the biased sampling, we deterministically select the natural frequencies of the oscillators, similarly to \cite{PhysRevE.81.056204}, as the $N$-tuple satisfying the constraints:
\begin{align*}
	\int_{-\infty}^{\omega_1}g(\omega) d\omega &=\frac{1}{N+1},  \quad (i=1)\\
	\int_{-\omega_{i-1}}^{\omega_i}g(\omega) d\omega &=\frac{1}{N+1}, \quad (i=2,\dots, N)
\end{align*}
where $g(\omega)$ is the probability density function of a given distribution. It should be noted that for a large network, we performed our simulation taking the natural frequencies randomly from a distribution and the obtained results are qualitatively the same.

Next, we investigate how the evolution of the network is affected by tuning the heterogeneity in the nodes. To this aim we use the edge snapping strategy described above in a novel evolutionary manner (see Fig. \ref{fig:1}) as explained in the next section.

\section{Evolutionary Edge-Snapping}
The evolutionary Edge-Snapping technique is based on two fundamental steps: one implementing the {\em variation ingredient} of evolution, the other its {\em selection mechanism}.

To implement the {\em variation ingredient} of evolution, a set of unweighted networks is generated using equations (\ref{equ:genphoscill}) and (\ref{equ:genedgesnph})  starting the process from different sets of initial conditions.
We consider a set of $n_S$ initial conditions randomly selected using a Latin Hypercube strategy \cite{McBe:79} in the range $\theta_n(0) \in [0, 2\pi[$, $n=1,2,\dots, N$.
To obtain the ``fitness'' of each link, 
we next compute the probability $p_{ij}$ of each link being activated as the fraction between the number of generated networks where that link is present, say $n_{ij}$, and the total number of trials, e.g. $p_{ij}= n_{ij} / n_S$. This yields a stochastic $N \times N$ matrix $P$ whose elements are the probabilities of activation of every possible link among nodes.

The {\em selection rule} is  obtained by selecting only those links whose activation probability is above a certain critical threshold value $p^*$, i.e. such that $p_{ij}>p^*$. We choose $p^*$ so as to guarantee that the resulting network is connected and has the smallest number of links. We shall term such a network as the {\em minimal edge-snapping (ES) network}.


\begin{figure*}[ht]
	\center
	\includegraphics{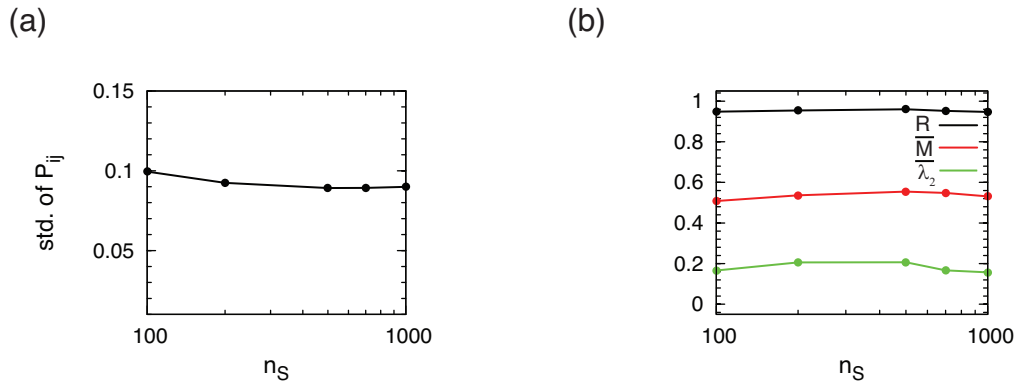}
	\caption{(a) Standard deviation of the link activation probability $p_{ij}$ as a function of the number of trials $n_S$. (b) Order parameter and relative number of links of the minimal ES structure as a function of the number of trials $n_S$.}
	\label{fig:A4} 
\end{figure*}
The variation step of our evolutionary strategy relies on the generation of a set of $n_S$ unweighted network using equations \eqref{equ:genphoscill} and \eqref{equ:genedgesnph} and starting the process from a different set of initial condition. With the aim of choosing a reasonable value for the number of trials $n_S$, we plot in Fig. \ref{fig:A4}(a) the standard deviation of the link activation probability $p_{ij}$ as a function of $n_S$. As can be noted, the differentiation in the $p_{ij}$ is quite constant as $n_S$ varies from $100$ to $1000$. Thus we select $n_S=100$ in all of our simulations. Indeed this guarantees a good degree of variation with the least computational cost. Finally, Fig. \ref{fig:A4}(b) confirms that the dynamical and structural properties of the emerging ES minimal structure do not show  significant fluctuations when the value of $n_S$ is increased. 

Note that the state space of the initial phases of many oscillators is a high-dimensional space (i.e. the aggregate N-dimensional state space obtained collecting the phase of each oscillator in the stack vector $\Theta=\left [ \theta_1, \theta_2, \dots, \theta_N \right ]$). 
To obtain effective samplings from that space, we adopted a Latin Hypercube Sampling (LHS) strategy first proposed in \cite{McBe:79}. LHS is a statistical method for generating a sample of plausible collections of parameter values from a multidimensional distribution. Specifically, let $X$ denote a $N$ variate random variable with probability density function $f(x)$ for $x \in S$. Then the range space of each of the $N$ components of $X$ is partitioned in $n_S$ disjoint intervals $S_i$ of size $p_i=P(X \in S_i)=1/n_S$. Taking the Cartesian product of these intervals yields $n_S^N$ cells each of probability size $n_S^{-N}$. Each cell can be labeled by a set of $N$ coordinates $m_i=(m_{i1}, m_{i,2}, \dots, m_{iN})$ where $m_{ij}$ is the interval number of component $X_j$ represented in cell $i$. A LHS is obtained from a random selection of the cells $m_1, \dots, m_{n_S}$, with the condition that for each $j$ the set $\{m_{ij}\}_{i=1}^{n_S}$ is a permutation of integers $1, 2, \dots, n_S$. As a result, one random observation is made in each cell. The main advantage of the LHS strategy is that it does not require more samples for more dimension of the range space $S$. This is the main reason why we use LHS in our method.   

To measure the synchronization performance of a ES network, we consider an ensemble of  phase oscillators connected by that  network and evaluate Kuramoto order parameter as
$R e^{i \psi} = \frac{1}{N}\sum_{n=1}^N e^{i \theta_n}$.   

\section{Emergence of Minimal networks}
\begin{figure}[t]
	\includegraphics{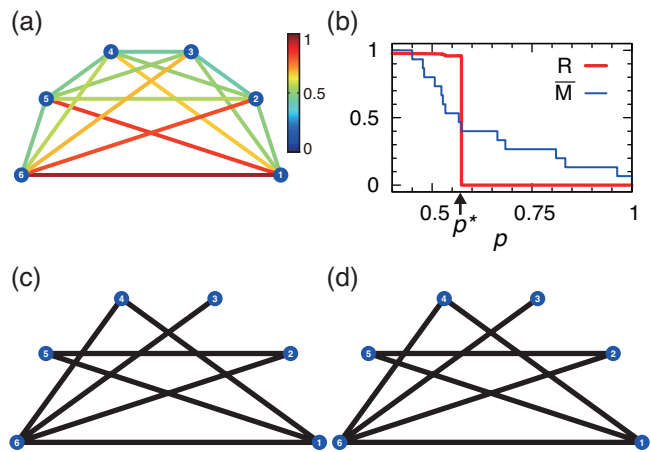}
	\caption{(a) Link activation probabilities $p_{ij}$ in the case of $N=6$ generated by the variation stage of the evolutionary ES strategy; 
		(b) Selection of the threshold probability value $p$: order parameter $R$, relative number of links ${\bar M}$.
		The arrow on the x-axis indicates the critical threshold $p^*$ which gives the minimal ES network;
		(c) Minimal Edge-Snapping Network; (d) Optimal network maximizing $R$ obtained by Exhaustive search and a Monte Carlo based method.}
	\label{fig:2}
\end{figure}

We first test our strategy by applying it to a small size network with $N=6$ and $\sigma = 0.3$  (Fig. \ref{fig:2}). We obtain the $P$ matrix visualized in Fig. 2(a). In Fig. 2(b), 
as the threshold value $p$ is increased, the number of edges, $M$, rapidly decreases while the value of the order parameter $R$ remains near unity.

\begin{figure*}
	\includegraphics{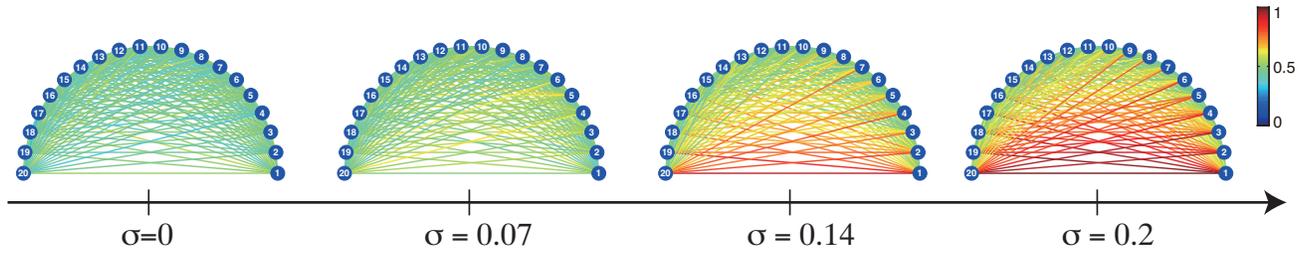}
	\caption{
		Heterogeneity induces functional structural properties of the network.
		$P$ matrix as a function of the heterogeneity parameter $\sigma$ when $N=20$. }
	\label{fig:3}
\end{figure*}
\begin{figure*}
	\includegraphics{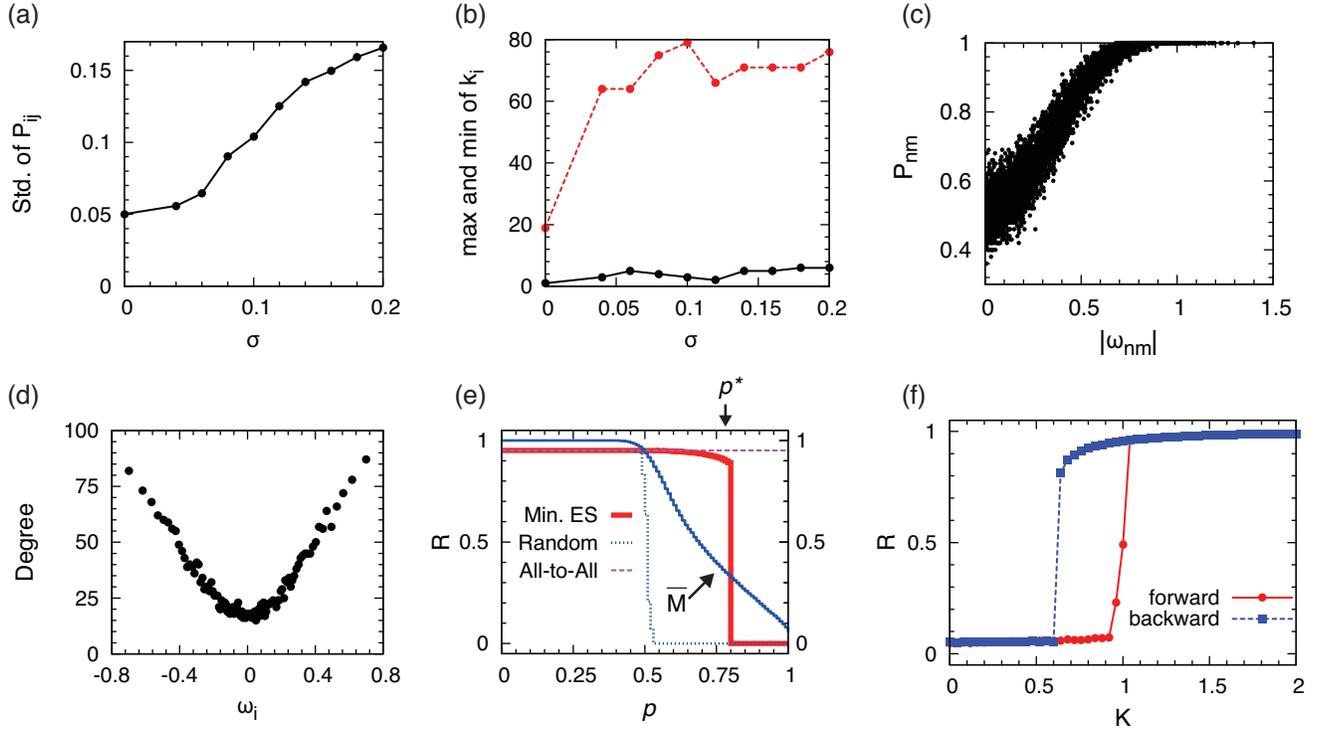}
	\caption{
		Structural properties of the emergent minimal ES network with $N=100$ and $\sigma=0.2$.
		(a) Standard deviation of the link activation probabilities $p_{ij}$ as a function of $\sigma$.
		(b) Maximum (red dashed line) and minimum (black solid line) value of Node Degree $k_i$ as a function of $\sigma$.
		(c) Activation probability of each link against the value of the difference between the natural frequencies of the oscillators at the endpoints.
		(d) Node degree $k_i$ vs. $\omega_i$.
		(e) Order parameter $R$ (red solid line) and relative number of links ${\bar M}$ (blue solid line) of the ES network as a function of the threshold probability value $p$. For comparison, the value $R$ is depicted for an all-to-all network (purple dashed line) and for randomly generated networks (blue dot-dashed line) with the same number of links.  The arrow on  x-axis represents the threshold $p^*$ to give the minimal ES network.
		(f) Order parameter $R$ of the phase oscillators interconnected by the minimal ES network when the overall coupling strength $K$ is increased (red solid) and decreased (blue dashed). We set $N=300$ and $\sigma=0.2$.
	}
	\label{fig:4}
\end{figure*}
In the figure, the normalized number of edges, which is divided by maximum links between $N$ nodes, i.e. ${\bar M}=M/M^{a2a}$, is plotted.
Also above a certain threshold the network becomes disconnected. Therefore we choose $p^*=0.57$ obtaining the minimal ES network depicted in Fig. 2(c) which is characterized by $M=7$ edges and $R=0.96$.
We compare the minimal ES structure  with the optimal  network structure shown in Fig. 2(d) obtained from an exhaustive search and a Monte Carlo based method \cite{PhysRevE.81.056212} maximizing the value of $R$ with the constraint that the total number of edges $M$ is equal to 7. 
We notice that the two networks share the same links. 

Next, we study how heterogeneity induces functional structural properties of the network. Figure \ref{fig:3} shows the $P$ matrix as a function of the heterogeneity parameter $\sigma$ when $N=20$. We see that as $\sigma$ is increased a differentiation becomes more and more apparent in the distribution of the link activation probabilities $p_{ij}$ with edges between oscillators with relatively different frequencies becoming more likely to occur in the minimal ES structure.

Fig. \ref{fig:4}(a) shows the standard deviation of the link activation probabilities $p_{ij}$ as a linear function of $\sigma$ in a larger network of $N=100$ oscillators. The structural properties of the emerging network are therefore induced by the node heterogeneity. This is confirmed in Fig. \ref{fig:4}(b) where the maximum and minimum values of the node degree $k_i$, corresponding to each minimal ES network, is plotted as a function of $\sigma$. The behaviors of the maximum value of $k_i$ (red dashed line) and the minimum of $k_i$ (black solid line) show an abrupt transition when passing from $\sigma=0$ to $\sigma>0$. This suggests that the differentiation in the degree distribution of the minimal ES network becomes remarkable when heterogeneity in the nodes is increased from zero (identical oscillators) to a value greater than zero (non-identical oscillators).

The structural properties of the emergent minimal ES network are highlighted in Fig. \ref{fig:4}(c)-(f) for a network of highly heterogeneous $N=100$ oscillators ($\sigma=0.2$). The activation probability of each link is plotted in Fig. \ref{fig:4}(c) against the value of the difference between the natural frequencies of the oscillators at the endpoints. Links connecting more distant nodes tend to be activated with a higher likelihood confirming that differentiation among links is induced by heterogeneity in the nodes. Also, as shown in Fig. \ref{fig:4}(d), hubs tend to be associated with oscillators whose frequencies are farther away from the average. 
The functional advantage of the emerging network is shown in Fig. \ref{fig:4}(e). Indeed, we observe that  the order parameter of the minimal ES structure is close to its maximal value for an all-to-all network of the same size, even if the number of links in the minimal ES network is remarkably lower than that in an all-to-all configuration.   
For the sake of comparison, the values of $R$  for a randomly generated network of the same number of edges is also depicted in Fig. \ref{fig:4}(e). 
The sudden dip of $R$ is due to the graph becoming disconnected beyond that critical value of the threshold $p^*$.

Notice that, as shown in Fig. \ref{fig:4}(f), 
as the coupling strength $K$ is varied, the order parameter $R$ of the phase oscillators interconnected by the  minimal ES network
exhibits a sudden hysteretic change, associated to a discontinuous phase transition,
whereas the system with a unimodal frequency distribution undergoes a continuous phase transition \cite{KuramotoChemicalOsci}.
This discontinuous phase transition, also known as ``explosive synchronisation'', has been studied in the literature \cite{pazo2005,GaGo:11,Leyva:2013aa,Zhang2013}, also in the case of adaptive networks \cite{HuHa:11,PhysRevLett.114.038701}, revealing that  the correlation between natural frequencies and the node degree, as shown in Fig. 4(d), can induce this phenomenon.
Here, we wish to emphasise that the proposed evolutionary strategy, which functionally organizes the network structure for synchronization, changes the type of phase transition that would be generically observed otherwise, inducing explosive synchronisation.

Our results clearly show the role of node heterogeneity in inducing functional structures using an evolutionary strategy for network synchronization. In particular, differences in the node dynamics do influence the evolution of the network determining a differentiation in the link activation probabilities that is instrumental to obtain minimal structures with relatively high values of the order parameter. Also, hubs tend to emerge there where the distance from the average natural frequency is highest. 
Further simulations also confirmed that a similar structure of the emergent network can be induced by using a power-law rather than a normal distribution when selecting the heterogeneous natural frequencies of the oscillators (data not shown).

It is notable that the presence of hubs seems to characterize the emergent networks for synchronization when the nodes are heterogeneous as opposed to more homogenous structures, such as entangled networks, which have been suggested to be optimal structures in the homogeneous case \cite{PhysRevLett.95.188701}. This is also confirmed in the case of Monte Carlo based optimal networks in \cite{PhysRevE.81.056204} where the presence of links between nodes with more distant frequencies is shown to be more likely and in the recent paper \cite{PhysRevLett.113.144101} based on the use of gradient-based methods. Here we obtain a further confirmation of these observations but via a generic local evolutionary strategy that is state-dependent and can be applied to a wider range of network synchronization and control problems. \\

\section{Conclusions}
Our results suggest that heterogeneity is the driving force determining the evolution of state-dependent functional networks. This can explain the structural properties  detected in natural networks such as neural interconnections in the brain, gene regulatory networks or  ecological networks where the states of the nodes typically affects the evolution of their interconnections \cite{Gross,CPLX:CPLX20386,PhysRevE.86.015101,Belykh20141}. It can also be used in Dynamical Systems and Control theory to design state-dependent evolutionary strategies able to induce a desired collective behavior in a network of interest.

\section*{ACKNOWLEDGMENTS}
This work was partially supported by JSPS KAKENHI Grants No. 24120708, No. 24740266, No. 25115719, and No. 26520206. FS would like to  acknowledge support from the Network of Excellence MASTRI Materiali e Strutture Intelligenti (POR Campania FSE 2007/2013).

\begin{thebibliography}{27}
	\expandafter\ifx\csname natexlab\endcsname\relax\def\natexlab#1{#1}\fi
	\expandafter\ifx\csname bibnamefont\endcsname\relax
	\def\bibnamefont#1{#1}\fi
	\expandafter\ifx\csname bibfnamefont\endcsname\relax
	\def\bibfnamefont#1{#1}\fi
	\expandafter\ifx\csname citenamefont\endcsname\relax
	\def\citenamefont#1{#1}\fi
	\expandafter\ifx\csname url\endcsname\relax
	\def\url#1{\texttt{#1}}\fi
	\expandafter\ifx\csname urlprefix\endcsname\relax\def\urlprefix{URL }\fi
	\providecommand{\bibinfo}[2]{#2}
	\providecommand{\eprint}[2][]{\url{#2}}
	
	\bibitem[{\citenamefont{Darwin}(1951)}]{Darw:51}
	\bibinfo{author}{\bibfnamefont{C.}~\bibnamefont{Darwin}},
	\emph{\bibinfo{title}{The Origin of Species by Means of Natural Selection, or
			the Preservation of Favoured Races in the Struggle for Life}}
	(\bibinfo{publisher}{Oxford Univ. Press}, \bibinfo{year}{1951}).
	
	\bibitem[{\citenamefont{Tan et~al.}(2014)\citenamefont{Tan, Lu, Chen, and
			Hill}}]{TaLu:14}
	\bibinfo{author}{\bibfnamefont{S.}~\bibnamefont{Tan}},
	\bibinfo{author}{\bibfnamefont{J.}~\bibnamefont{Lu}},
	\bibinfo{author}{\bibfnamefont{G.}~\bibnamefont{Chen}}, \bibnamefont{and}
	\bibinfo{author}{\bibfnamefont{D.}~\bibnamefont{Hill}},
	\bibinfo{journal}{Circuits and Systems Magazine, IEEE}
	\textbf{\bibinfo{volume}{14}}, \bibinfo{pages}{36} (\bibinfo{year}{2014}).
	
	\bibitem[{\citenamefont{Boccaletti et~al.}(2006)\citenamefont{Boccaletti,
			Latora, Moreno, Chavez, and Hwang}}]{Boccaletti2006175}
	\bibinfo{author}{\bibfnamefont{S.}~\bibnamefont{Boccaletti}},
	\bibinfo{author}{\bibfnamefont{V.}~\bibnamefont{Latora}},
	\bibinfo{author}{\bibfnamefont{Y.}~\bibnamefont{Moreno}},
	\bibinfo{author}{\bibfnamefont{M.}~\bibnamefont{Chavez}}, \bibnamefont{and}
	\bibinfo{author}{\bibfnamefont{D.-U.} \bibnamefont{Hwang}},
	\bibinfo{journal}{Physics Reports} \textbf{\bibinfo{volume}{424}},
	\bibinfo{pages}{175 } (\bibinfo{year}{2006}).
	
	\bibitem[{\citenamefont{Barrat et~al.}(2008)\citenamefont{Barrat,
			Barth{\'e}lemy, and Vespignani}}]{barrat2008dynamical}
	\bibinfo{author}{\bibfnamefont{A.}~\bibnamefont{Barrat}},
	\bibinfo{author}{\bibfnamefont{M.}~\bibnamefont{Barth{\'e}lemy}},
	\bibnamefont{and}
	\bibinfo{author}{\bibfnamefont{A.}~\bibnamefont{Vespignani}},
	\emph{\bibinfo{title}{{Dynamical Processes on Complex Networks}}}
	(\bibinfo{publisher}{Cambridge University Press}, \bibinfo{year}{2008}).
	
	\bibitem[{\citenamefont{Kuramoto}(1984)}]{KuramotoChemicalOsci}
	\bibinfo{author}{\bibfnamefont{Y.}~\bibnamefont{Kuramoto}},
	\emph{\bibinfo{title}{Chemical Oscillations, Waves, and Turbulence}}
	(\bibinfo{publisher}{Springer, New York}, \bibinfo{year}{1984}).
	
	\bibitem[{\citenamefont{Strogatz}(2000)}]{Strogatz20001}
	\bibinfo{author}{\bibfnamefont{S.~H.} \bibnamefont{Strogatz}},
	\bibinfo{journal}{Physica D: Nonlinear Phenomena}
	\textbf{\bibinfo{volume}{143}}, \bibinfo{pages}{1 } (\bibinfo{year}{2000}).
	
	\bibitem[{\citenamefont{Acebr\'on et~al.}(2005)\citenamefont{Acebr\'on,
			Bonilla, P\'erez~Vicente, Ritort, and Spigler}}]{KuramotoReview}
	\bibinfo{author}{\bibfnamefont{J.~A.} \bibnamefont{Acebr\'on}},
	\bibinfo{author}{\bibfnamefont{L.~L.} \bibnamefont{Bonilla}},
	\bibinfo{author}{\bibfnamefont{C.~J.} \bibnamefont{P\'erez~Vicente}},
	\bibinfo{author}{\bibfnamefont{F.}~\bibnamefont{Ritort}}, \bibnamefont{and}
	\bibinfo{author}{\bibfnamefont{R.}~\bibnamefont{Spigler}},
	\bibinfo{journal}{Rev. Mod. Phys.} \textbf{\bibinfo{volume}{77}},
	\bibinfo{pages}{137} (\bibinfo{year}{2005}).
	
	\bibitem[{\citenamefont{Arenas et~al.}(2008)\citenamefont{Arenas,
			D{\'\i}az-Guilera, Kurths, Moreno, and Zhou}}]{ArGu:08}
	\bibinfo{author}{\bibfnamefont{A.}~\bibnamefont{Arenas}},
	\bibinfo{author}{\bibfnamefont{A.}~\bibnamefont{D{\'\i}az-Guilera}},
	\bibinfo{author}{\bibfnamefont{J.}~\bibnamefont{Kurths}},
	\bibinfo{author}{\bibfnamefont{Y.}~\bibnamefont{Moreno}}, \bibnamefont{and}
	\bibinfo{author}{\bibfnamefont{C.}~\bibnamefont{Zhou}},
	\bibinfo{journal}{Physics Reports} \textbf{\bibinfo{volume}{469}},
	\bibinfo{pages}{93 } (\bibinfo{year}{2008}).
	
	\bibitem[{\citenamefont{Donetti et~al.}(2005)\citenamefont{Donetti, Hurtado,
			and Mu\~noz}}]{PhysRevLett.95.188701}
	\bibinfo{author}{\bibfnamefont{L.}~\bibnamefont{Donetti}},
	\bibinfo{author}{\bibfnamefont{P.~I.} \bibnamefont{Hurtado}},
	\bibnamefont{and} \bibinfo{author}{\bibfnamefont{M.~A.}
		\bibnamefont{Mu\~noz}}, \bibinfo{journal}{Phys. Rev. Lett.}
	\textbf{\bibinfo{volume}{95}}, \bibinfo{pages}{188701}
	(\bibinfo{year}{2005}).
	
	\bibitem[{\citenamefont{Rad et~al.}(2008)\citenamefont{Rad, Jalili, and
			Hasler}}]{Chaos2008}
	\bibinfo{author}{\bibfnamefont{A.~A.} \bibnamefont{Rad}},
	\bibinfo{author}{\bibfnamefont{M.}~\bibnamefont{Jalili}}, \bibnamefont{and}
	\bibinfo{author}{\bibfnamefont{M.}~\bibnamefont{Hasler}},
	\bibinfo{journal}{Chaos: An Interdisciplinary Journal of Nonlinear Science}
	\textbf{\bibinfo{volume}{18}},  (\bibinfo{year}{2008}).
	
	\bibitem[{\citenamefont{Yanagita and Mikhailov}(2010)}]{PhysRevE.81.056204}
	\bibinfo{author}{\bibfnamefont{T.}~\bibnamefont{Yanagita}} \bibnamefont{and}
	\bibinfo{author}{\bibfnamefont{A.~S.} \bibnamefont{Mikhailov}},
	\bibinfo{journal}{Phys. Rev. E} \textbf{\bibinfo{volume}{81}},
	\bibinfo{pages}{056204} (\bibinfo{year}{2010}).
	
	\bibitem[{\citenamefont{Gorochowski et~al.}(2010)\citenamefont{Gorochowski,
			di~Bernardo, and Grierson}}]{PhysRevE.81.056212}
	\bibinfo{author}{\bibfnamefont{T.~E.} \bibnamefont{Gorochowski}},
	\bibinfo{author}{\bibfnamefont{M.}~\bibnamefont{di~Bernardo}},
	\bibnamefont{and} \bibinfo{author}{\bibfnamefont{C.~S.}
		\bibnamefont{Grierson}}, \bibinfo{journal}{Phys. Rev. E}
	\textbf{\bibinfo{volume}{81}}, \bibinfo{pages}{056212}
	(\bibinfo{year}{2010}).
	
	\bibitem[{\citenamefont{Tanaka and Aoyagi}(2008)}]{Tanaka2008b}
	\bibinfo{author}{\bibfnamefont{T.}~\bibnamefont{Tanaka}} \bibnamefont{and}
	\bibinfo{author}{\bibfnamefont{T.}~\bibnamefont{Aoyagi}},
	\bibinfo{journal}{Physical Review E} \textbf{\bibinfo{volume}{78}},
	\bibinfo{pages}{046210} (\bibinfo{year}{2008}).
	
	\bibitem[{\citenamefont{Skardal et~al.}(2014)\citenamefont{Skardal, Taylor, and
			Sun}}]{PhysRevLett.113.144101}
	\bibinfo{author}{\bibfnamefont{P.~S.} \bibnamefont{Skardal}},
	\bibinfo{author}{\bibfnamefont{D.}~\bibnamefont{Taylor}}, \bibnamefont{and}
	\bibinfo{author}{\bibfnamefont{J.}~\bibnamefont{Sun}},
	\bibinfo{journal}{Phys. Rev. Lett.} \textbf{\bibinfo{volume}{113}},
	\bibinfo{pages}{144101} (\bibinfo{year}{2014}).
	
	\bibitem[{\citenamefont{Guti\'errez et~al.}(2011)\citenamefont{Guti\'errez,
			Amann, Assenza, G\'omez-Garde\~nes, Latora, and Boccaletti}}]{GuAm:11}
	\bibinfo{author}{\bibfnamefont{R.}~\bibnamefont{Guti\'errez}},
	\bibinfo{author}{\bibfnamefont{A.}~\bibnamefont{Amann}},
	\bibinfo{author}{\bibfnamefont{S.}~\bibnamefont{Assenza}},
	\bibinfo{author}{\bibfnamefont{J.}~\bibnamefont{G\'omez-Garde\~nes}},
	\bibinfo{author}{\bibfnamefont{V.}~\bibnamefont{Latora}}, \bibnamefont{and}
	\bibinfo{author}{\bibfnamefont{S.}~\bibnamefont{Boccaletti}},
	\bibinfo{journal}{Phys. Rev. Lett.} \textbf{\bibinfo{volume}{107}}
	(\bibinfo{year}{2011}).
	
	\bibitem[{\citenamefont{DeLellis et~al.}(2010)\citenamefont{DeLellis,
			di~Bernardo, Garofalo, and Porfiri}}]{EdgeSnapping}
	\bibinfo{author}{\bibfnamefont{P.}~\bibnamefont{DeLellis}},
	\bibinfo{author}{\bibfnamefont{M.}~\bibnamefont{di~Bernardo}},
	\bibinfo{author}{\bibfnamefont{F.}~\bibnamefont{Garofalo}}, \bibnamefont{and}
	\bibinfo{author}{\bibfnamefont{M.}~\bibnamefont{Porfiri}},
	\bibinfo{journal}{Circuits and Systems I: Regular Papers, IEEE Transactions
		on} \textbf{\bibinfo{volume}{57}}, \bibinfo{pages}{2132}
	(\bibinfo{year}{2010}).
	
	\bibitem[{\citenamefont{McKay et~al.}(1979)\citenamefont{McKay, Beckman, and
			Conover}}]{McBe:79}
	\bibinfo{author}{\bibfnamefont{M.~D.} \bibnamefont{McKay}},
	\bibinfo{author}{\bibfnamefont{R.~J.} \bibnamefont{Beckman}},
	\bibnamefont{and} \bibinfo{author}{\bibfnamefont{W.~J.}
		\bibnamefont{Conover}}, \bibinfo{journal}{Technometrics}
	\textbf{\bibinfo{volume}{21}}, \bibinfo{pages}{239} (\bibinfo{year}{1979}).
	
	\bibitem[{\citenamefont{Paz\'o}(2005)}]{pazo2005}
	\bibinfo{author}{\bibfnamefont{D.}~\bibnamefont{Paz\'o}},
	\bibinfo{journal}{Phys. Rev. E} \textbf{\bibinfo{volume}{72}},
	\bibinfo{pages}{046211} (\bibinfo{year}{2005}).
	
	\bibitem[{\citenamefont{G\'omez-Garde\~nes
			et~al.}(2011)\citenamefont{G\'omez-Garde\~nes, G\'omez, Arenas, and
			Moreno}}]{GaGo:11}
	\bibinfo{author}{\bibfnamefont{J.}~\bibnamefont{G\'omez-Garde\~nes}},
	\bibinfo{author}{\bibfnamefont{S.}~\bibnamefont{G\'omez}},
	\bibinfo{author}{\bibfnamefont{A.}~\bibnamefont{Arenas}}, \bibnamefont{and}
	\bibinfo{author}{\bibfnamefont{Y.}~\bibnamefont{Moreno}},
	\bibinfo{journal}{Phys. Rev. Lett.} \textbf{\bibinfo{volume}{106}},
	\bibinfo{pages}{128701} (\bibinfo{year}{2011}).
	
	\bibitem[{\citenamefont{Leyva et~al.}(2013)\citenamefont{Leyva, Navas,
			Sendi{\~n}a-Nadal, Almendral, Buld{\'u}, Zanin, Papo, and
			Boccaletti}}]{Leyva:2013aa}
	\bibinfo{author}{\bibfnamefont{I.}~\bibnamefont{Leyva}},
	\bibinfo{author}{\bibfnamefont{A.}~\bibnamefont{Navas}},
	\bibinfo{author}{\bibfnamefont{I.}~\bibnamefont{Sendi{\~n}a-Nadal}},
	\bibinfo{author}{\bibfnamefont{J.~A.} \bibnamefont{Almendral}},
	\bibinfo{author}{\bibfnamefont{J.~M.} \bibnamefont{Buld{\'u}}},
	\bibinfo{author}{\bibfnamefont{M.}~\bibnamefont{Zanin}},
	\bibinfo{author}{\bibfnamefont{D.}~\bibnamefont{Papo}}, \bibnamefont{and}
	\bibinfo{author}{\bibfnamefont{S.}~\bibnamefont{Boccaletti}},
	\bibinfo{journal}{Sci. Rep.} \textbf{\bibinfo{volume}{3}}
	(\bibinfo{year}{2013}).
	
	\bibitem[{\citenamefont{Zhang et~al.}(2013)\citenamefont{Zhang, Hu, Kurths, and
			Liu}}]{Zhang2013}
	\bibinfo{author}{\bibfnamefont{X.}~\bibnamefont{Zhang}},
	\bibinfo{author}{\bibfnamefont{X.}~\bibnamefont{Hu}},
	\bibinfo{author}{\bibfnamefont{J.}~\bibnamefont{Kurths}}, \bibnamefont{and}
	\bibinfo{author}{\bibfnamefont{Z.}~\bibnamefont{Liu}},
	\bibinfo{journal}{Phys. Rev. E} \textbf{\bibinfo{volume}{88}},
	\bibinfo{pages}{010802} (\bibinfo{year}{2013}).
	
	\bibitem[{\citenamefont{Hui-Jun et~al.}(2011)\citenamefont{Hui-Jun, Hao, and
			Zhong-Huai}}]{HuHa:11}
	\bibinfo{author}{\bibfnamefont{J.}~\bibnamefont{Hui-Jun}},
	\bibinfo{author}{\bibfnamefont{W.}~\bibnamefont{Hao}}, \bibnamefont{and}
	\bibinfo{author}{\bibfnamefont{H.}~\bibnamefont{Zhong-Huai}},
	\bibinfo{journal}{Chinese Physics Letters} \textbf{\bibinfo{volume}{28}}
	(\bibinfo{year}{2011}).
	
	\bibitem[{\citenamefont{Zhang et~al.}(2015)\citenamefont{Zhang, Boccaletti,
			Guan, and Liu}}]{PhysRevLett.114.038701}
	\bibinfo{author}{\bibfnamefont{X.}~\bibnamefont{Zhang}},
	\bibinfo{author}{\bibfnamefont{S.}~\bibnamefont{Boccaletti}},
	\bibinfo{author}{\bibfnamefont{S.}~\bibnamefont{Guan}}, \bibnamefont{and}
	\bibinfo{author}{\bibfnamefont{Z.}~\bibnamefont{Liu}},
	\bibinfo{journal}{Phys. Rev. Lett.} \textbf{\bibinfo{volume}{114}},
	\bibinfo{pages}{038701} (\bibinfo{year}{2015}).
	
	\bibitem[{\citenamefont{Gross and Blasius}(2008)}]{Gross}
	\bibinfo{author}{\bibfnamefont{T.}~\bibnamefont{Gross}} \bibnamefont{and}
	\bibinfo{author}{\bibfnamefont{B.}~\bibnamefont{Blasius}},
	\bibinfo{journal}{J. R. Soc. Interface} \textbf{\bibinfo{volume}{5}},
	\bibinfo{pages}{259} (\bibinfo{year}{2008}).
	
	\bibitem[{\citenamefont{Gorochowski et~al.}(2012)\citenamefont{Gorochowski,
			di~Bernardo, and Grierson}}]{CPLX:CPLX20386}
	\bibinfo{author}{\bibfnamefont{T.~E.} \bibnamefont{Gorochowski}},
	\bibinfo{author}{\bibfnamefont{M.}~\bibnamefont{di~Bernardo}},
	\bibnamefont{and} \bibinfo{author}{\bibfnamefont{C.~S.}
		\bibnamefont{Grierson}}, \bibinfo{journal}{Complexity}
	\textbf{\bibinfo{volume}{17}}, \bibinfo{pages}{18} (\bibinfo{year}{2012}).
	
	\bibitem[{\citenamefont{Avalos-Gayt\'an
			et~al.}(2012)\citenamefont{Avalos-Gayt\'an, Almendral, Papo, Schaeffer, and
			Boccaletti}}]{PhysRevE.86.015101}
	\bibinfo{author}{\bibfnamefont{V.}~\bibnamefont{Avalos-Gayt\'an}},
	\bibinfo{author}{\bibfnamefont{J.~A.} \bibnamefont{Almendral}},
	\bibinfo{author}{\bibfnamefont{D.}~\bibnamefont{Papo}},
	\bibinfo{author}{\bibfnamefont{S.~E.} \bibnamefont{Schaeffer}},
	\bibnamefont{and}
	\bibinfo{author}{\bibfnamefont{S.}~\bibnamefont{Boccaletti}},
	\bibinfo{journal}{Phys. Rev. E} \textbf{\bibinfo{volume}{86}},
	\bibinfo{pages}{015101} (\bibinfo{year}{2012}).
	
	\bibitem[{\citenamefont{Belykh et~al.}(2014)\citenamefont{Belykh, di~Bernardo,
			Kurths, and Porfiri}}]{Belykh20141}
	\bibinfo{author}{\bibfnamefont{I.}~\bibnamefont{Belykh}},
	\bibinfo{author}{\bibfnamefont{M.}~\bibnamefont{di~Bernardo}},
	\bibinfo{author}{\bibfnamefont{J.}~\bibnamefont{Kurths}}, \bibnamefont{and}
	\bibinfo{author}{\bibfnamefont{M.}~\bibnamefont{Porfiri}},
	\bibinfo{journal}{Physica D: Nonlinear Phenomena}
	\textbf{\bibinfo{volume}{267}}, \bibinfo{pages}{1 } (\bibinfo{year}{2014}).
	
\end{thebibliography}

\appendix
\section{DESCRIPTION OF THE EDGE SNAPPING METHOD }

\begin{figure*}[ht]
	\center
	\includegraphics{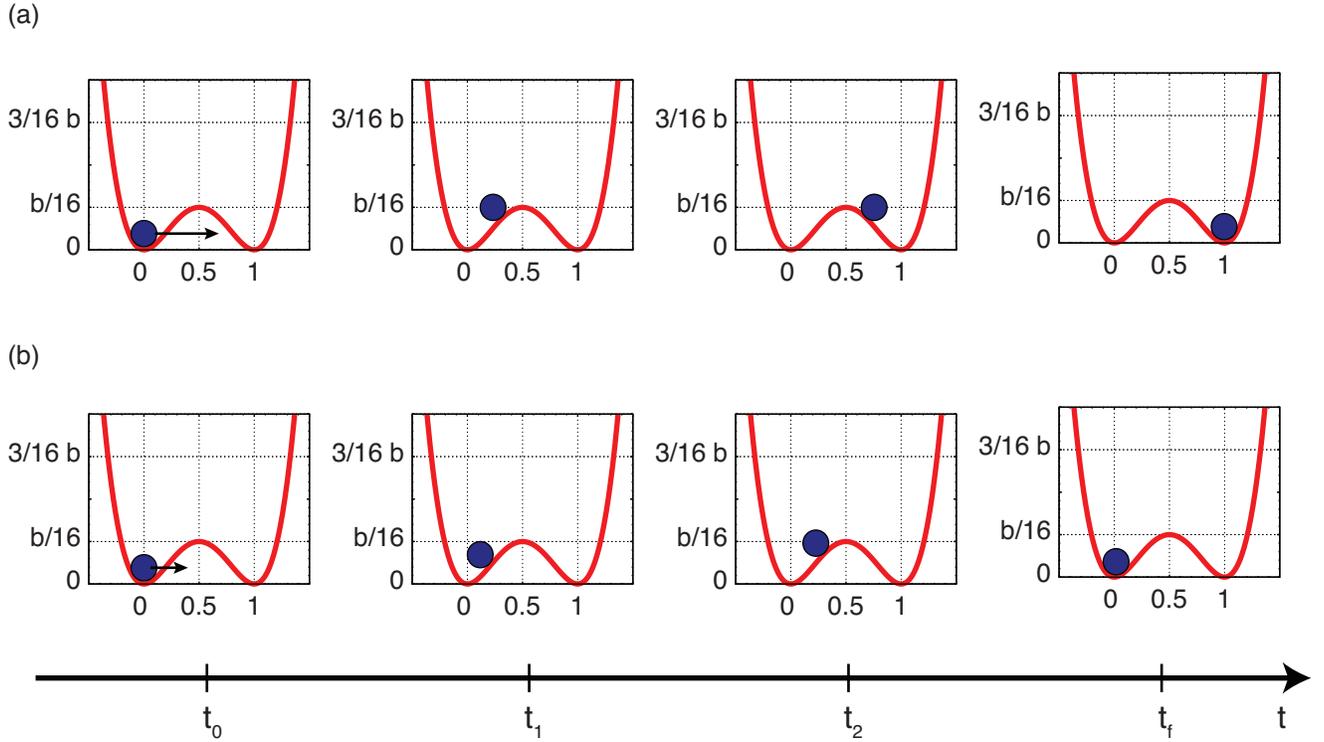}
	\caption{ Edges' evolution according to the edge snapping mechanism.}
	\label{fig:A1} 
\end{figure*}

\begin{figure*}[ht]
	\center
	\includegraphics{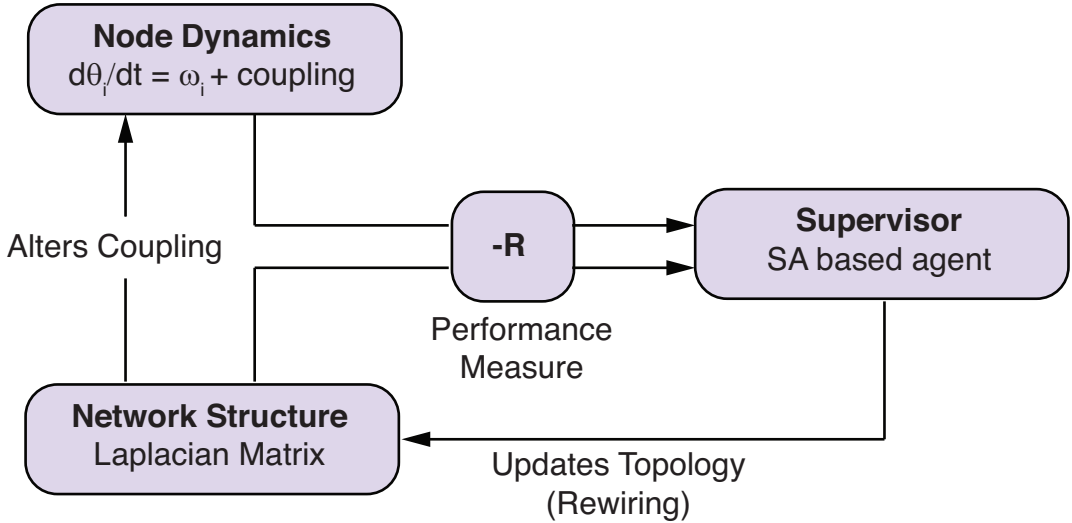}
	\caption{ Schematic illustration of how NetEvo works. }
	\label{fig:A2} 
\end{figure*}

Edge Snapping \cite{EdgeSnapping} is an adaptive strategy for the evolution of an unweighted network.
Time-varying coupling gains $k_{nm}$ are assigned to all pair of  nodes $n$ and $m$,
with a second order dynamics affected by a double well potential $V(k_{nm})=b k_{nm}^2(k_{nm}-1)^2$ defined as:
$$
\ddot{k}_{nm}+d \ \dot{k}_{nm}+\frac{\partial V(k_{nm})}{\partial k_{nm}}=h(\| \bm{x}_m- \bm{x}_n\|),
$$
where $d$ is a damping coefficient, $\bm{x}_n$ and $\bm{x}_m$ are the states of the nodes at the endpoints of the edge $(n, m)$.
The driving force $h(\| \bm{x}_m- \bm{x}_n\|)$ is a generic increasing function such that $h(0)=0$.

The gains' dynamics mimics the damped motion of a particle in a one-dimensional space subject to a double-well potential as schematically outlined in Fig. \ref{fig:A1}. 
Indeed, in Fig. \ref{fig:A1}(a) the initial forcing is strong enough to drive the mass particle from the equilibrium at 0 (edge turned off) to the well associated to the equilibrium at 1 (edge activated). On the contrary, in Fig. \ref{fig:A1}(b) the forcing input to the edge snapping dynamics is not able to move the particle from the equilibrium at the origin.
As a result of these dynamics, each coupling gains $k_{nm}$ converges to either one of the equilibrium points, $0$ or $1$. 

Note that the dynamics of the gains $k_{nm}$ is interdependent on the dynamics of the nodes' state $\bm{x}_n$ (in this letter, the dynamics of the states is given by a coupled oscillator dynamics among nodes).
The resulting unweighted network is the outcome of the co-evolving dynamics of the nodes and the state-dependent network.
This strategy is based on a distributed adaptive nonlinear approach and is therefore a generic decentralized approach relying only on a nonlinear potential to drive edge adaptation as explained in \cite{EdgeSnapping}.

In addition, this edge-snapping strategy has two parameters $b$ and $d$, which can be used to control  network evolution.
Indeed the barrier of the potential between the two wells, acts as a constraint.
As explained above, if the driving force is not strong enough, the edge, after a transient, will remain in the well corresponding to the absence of link.
The height of the barrier can be tuned varying the parameter $b$ in the expression of the potential $V$. The higher the barrier $b$, the stronger the constraint.

\section{NETEVO }

NetEvo is a computational framework designed to help understand the evolution of dynamical complex networks \cite{CPLX:CPLX20386}. It provides flexible tools for the simulation of dynamical processes on networks and methods for the evolution of underlying topological structures. To bring together simulation and evolution in a coherent way, the framework uses the idea of a supervisor, illustrated in Fig. \ref{fig:A2}. Evolution of the system is performed by the supervisor which can be viewed as a form of optimiser. This takes as input an initial topology, simulated output from the system and user defined constraints, and aims to return an optimal or enhanced topology. Changes to the system are assessed by using the performance measure -R (the opposite of the order parameter), with smaller values representing an improved performance. By default, NetEvo provides a supervisor that uses a Simulated Annealing meta-heuristic to search for near optimal configurations. This method has been shown to perform well for a wide range of problems with an unknown prior structure. 

We tuned NetEvo to find an optimal structure, given an initial condition (the same used in the procedure for finding the minimal structure). 
Simulated annealing tends to avoid local minima (or maxima), so we could start the optimization from any random connected network structure. However, we decided to start "near"  the minimal structure, to facilitate the optimization (by near, we mean a structure obtained from the minimal structure, after rewiring about  $10\%$ of its edges).

We note here, that it was necessary to run NetEvo several times (i.e. $n_{NE}=10$ times), because of local maxima traps that the algorithm could not avoid. Finally, the optimal (or sub-optimal) structure is selected as the network that maximize $R$ (starting from $\theta_0$), among each of the $n_{NE}$ results.

\end{document}